\documentclass{article}
\usepackage{bigsky2007}
\usepackage{graphicx}
\frompage{000} \topage{000}                                              
\def\rightmark{Identification of Bottom Contribution}
\def\leftmark{A. One et al.}

\newcommand{\dAu}{\textit{d}+Au}
\newcommand{\AuAu}{Au+Au}
\newcommand{\pp}{\mbox{\textit{p}+\textit{p}}}

\newcommand{\pt}{\mbox{$p_T$}}

\newcommand{\gev}{\mbox{$\mathrm{GeV}$}}

\newcommand{\gevc}{\mbox{${\mathrm{GeV/}}c$}}

\newcommand{\RAA}{\mbox{$R_{AA}$}}

\title{Identification of Bottom Contribution in Non-photonic Electron
Spectra and $v_2$ at RHIC}

\authors{
{Yifei Zhang$^1$ %
}\\[2.812mm]
{\normalsize
\hspace*{-8pt}$^1$ Dept. of Modern Physics, University of Science and Technology of China, Hefei, Anhui, China, 230026\\[0.2ex]
Lawrence Berkeley National Laboratory, 1 Cyclotron Road, MS70R319,
Berkeley, CA, 94720 \\
}}

\abstract{We present a study on the spectra for heavy flavor
(charm and bottom) decayed electrons in 200 \gev\ \pp\ collisions
and provide the relative contributions of charm and bottom hadrons
from the PYTHIA calculations. The results suggest that the
crossing point of the electron spectra from charm and bottom
decays is above 7 \gevc\ and bottom contribution is not dominant
for electron $\pt<3$ \gevc. The upper limit of the relative cross
section ratio is reported as
$\sigma_{b\bar{b}}/\sigma_{c\bar{c}}\leq(0.49\pm0.09\pm0.09)\%$.
We also compare the $v_2$ distribution from simulation to the
experimental data in 200 \gev\ \AuAu\ collisions and estimate the
possible charm $v_2$.}


\begin{document}

\maketitle
\setcounter{page}{1}

\section{Introduction}\label{intro}

Due to the absence of the measurement of B-mesons and precise
measurement of D-mesons, it is difficult to separate bottom and
charm contributions experimentally in current non-photonic
electron measurements for both spectra and elliptic flow $v_2$. As
discussed previously, the suppression behavior of heavy quarks is
quite different from light quarks due to the "dead cone"
effect~\cite{deadcone}, and this is especially true for the bottom
quark. Even when the elastic energy loss is included, the bottom
quark still loses much less energy. The bottom contribution may
reduce the energy loss of non-photonic electrons from heavy flavor
decays. But recent measurements show that the suppression of the
non-photonic electron \RAA\ is as large as light
hadrons~\cite{starcraa}. Both the theoretical result with charm
energy loss only and the theoretical calculations with
charm+bottom energy loss by assuming large $\hat{q}$ or counting
elastic energy loss can describe the data within
errors~\cite{DGLV06,Wicks05,RappRaa,Ivancoll,armeloss}.

Recently, PHENIX has measured the non-photonic electron
$v_2$~\cite{Phenixv2}. The observed large elliptic flow of the
non-photonic electron may indicate strong coupling of heavy quarks
with medium. There are many theoretical calculations for the
non-photonic electron $v_2$, such as charm thermal+flow
model~\cite{kocharmflow}, A multi-phase transition (AMPT) model
assume cross section $\sigma_{p}$=10 mb~\cite{AmptCharmflow},
resonance states of D-/B- mesons~\cite{vanHCharmflow}, {\em etc.}
The comparison with theories also showes that both the model
results with charm only and the results with charm+bottom have
good agreement with data within errors.

Thus, the puzzle of the bottom contributions in non-photonic
electron spectra and $v_2$ still remains. We present the following
method to estimate the bottom contributions and to study the
possible charm $v_2$.

\section{Fit to non-photonic electron spectrum and relative cross section ratio}\label{techno}

The non-photonic electron spectrum up to 10 \gevc\ has been
measured by STAR experiment in 200 \gev\ \pp\ collisions. The idea
is that we use the sum of electron spectra from both charm and
bottom decays in PYTHIA model~\cite{pythia} to fit the STAR \pp\
data~\cite{starcraa} to extract the fraction of the bottom
contribution. Since the D-mesons and their decay electrons spectra
from default PYTHIA parameters are soft~\cite{ffcharm}, a modified
Peterson Fragment Function (FF) and the high \pt\ tuned parameter
are used to make spectra harder to be comparable with the form
factor decays~\cite{XYLin04}.

Table~\ref{PYpar} lists the parameter initialization for PYTHIA
6.131:

\begin{table}[hbt]
\caption[PYTHIA parameters for heavy flavor decays]{PYTHIA
parameters for heavy flavor decays.} \label{PYpar} \vskip 0.1 in
\centering\begin{tabular}{|c|c|} \hline \hline
Parameter & Value \\
\hline
MSEL    & 4 (charm), 5 (bottom) \\
\hline
quark mass & $m_c=1.25$, $m_b=4.8$ (\gev) \\
\hline
parton dist. function & CTEQ5L \\
\hline
$Q^2$ scale & 4 \\
\hline
K factor & 3.5 \\
\hline
$\langle K_t\rangle$ & 1.5 \\
\hline
Peterson Frag. function & $\varepsilon=10^{-5}$ \\
\hline
high \pt\ tuned PARP(67) & 4 \\
     \hline \hline
\end{tabular}
\end{table}

\begin{figure} \centering \begin{minipage}[c]{0.47\textwidth} \centering
\includegraphics[width=1.\textwidth]{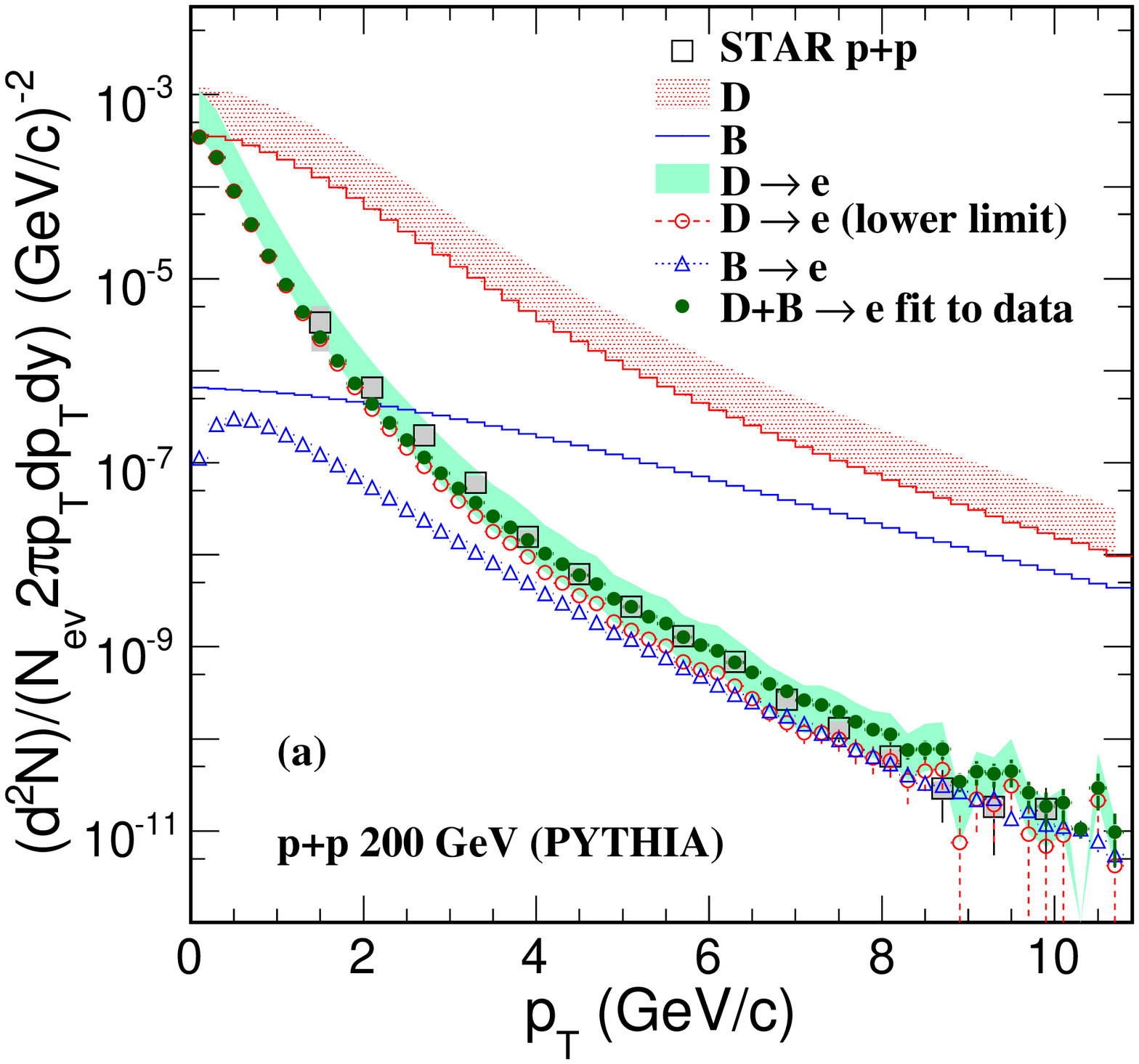}
\end{minipage}%
\begin{minipage}[c]{0.47\textwidth} \centering
\includegraphics[width=1.\textwidth]{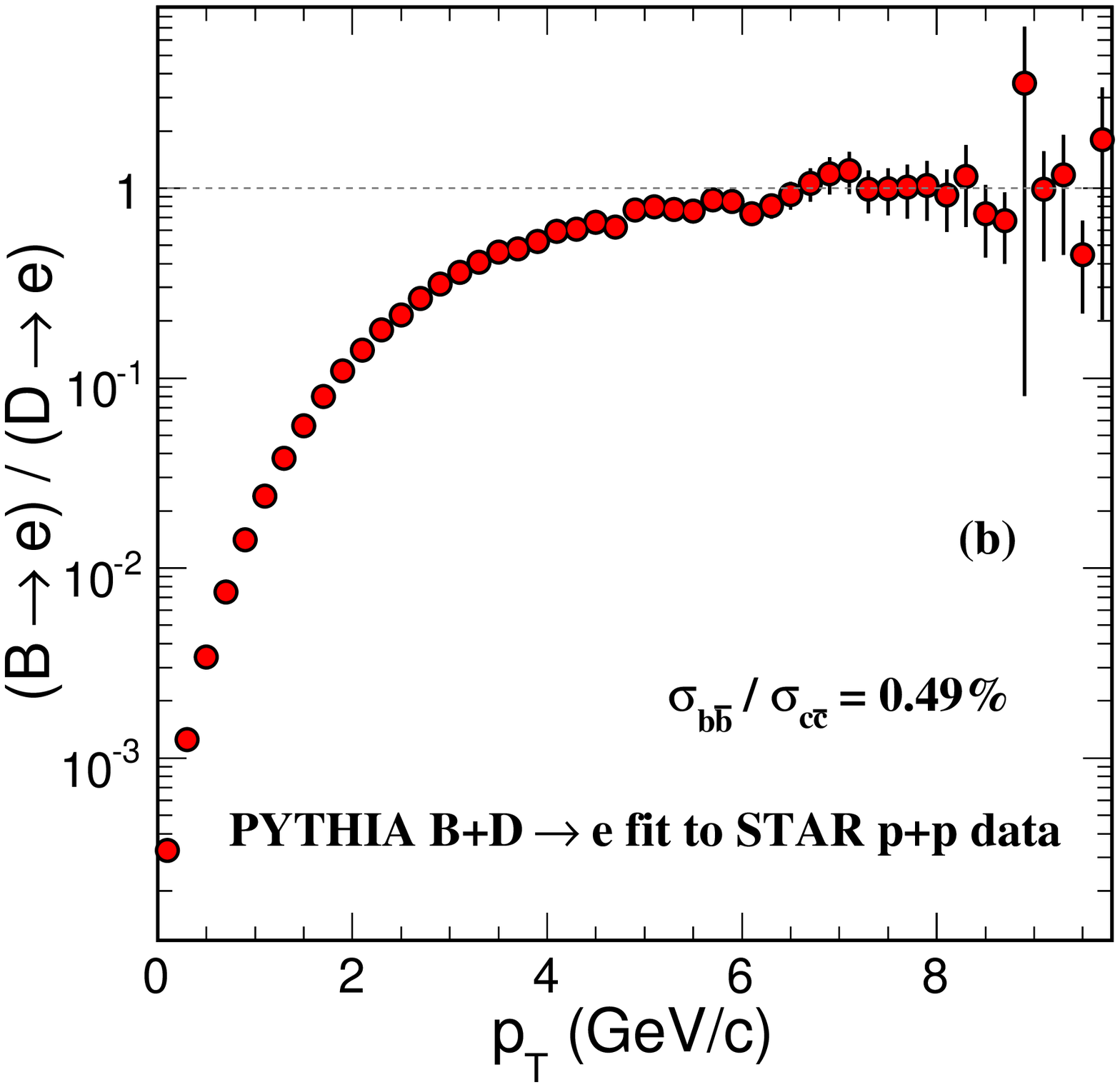}
\end{minipage}%
\caption[D-/B- mesons and their decay electron spectra from PYTHIA
and the relative spectra ratio]{Panel (a): D-/B- mesons and their
decay electron spectra from PYTHIA. The $B+D\rightarrow e$ fit to
STAR non-photonic electron data in \pp\ collisions. Panel (b): The
relative spectra ratio, upper limit of $B\rightarrow e$
contributions as a function of \pt.} \label{pyspecratio}
\end{figure}

Fig.~\ref{pyspecratio} (a) shows the \pt\ distributions of the
heavy flavor hadrons and their decay electrons from PYTHIA with
above parameters. The D-meson spectrum, shown as the hatched band,
is normalized to \begin{equation} dN/dy = dN/dy(D^0)/\langle
N_{bin}\rangle/R_{dAu}/R, \end{equation} where
$dN/dy(D^0)=0.028\pm0.004\pm0.008$ measured in \dAu\
collisions~\cite{stardAucharm}. $\langle N_{bin}\rangle=7.5\pm0.4$
in \dAu\ collisions. $R_{dAu}=1.3\pm0.3$~\cite{Xinthesis}. $R$
factor stands for $D^0$ fraction in total charmed hadrons, the
fragmentation ratio $R(c\rightarrow D^0)\equiv
N_{D^0}/N_{c\bar{c}}=0.54\pm0.05$~\cite{PDG}. All these
normalization errors are propagated into the uncertainty band of
the D-meson spectrum. The curve in this band is the lower limit of
the D-meson spectrum in our simulation. Correspondingly, its decay
electron spectrum is shown as the solid band. The non-photonic
electron spectrum measured in \pp\ collisions at
STAR~\cite{starcraa} is shown as the open squares. The decay
electron band alone can describe the data, indicating that the
contribution of electrons from bottom decay could be very small.
In order to estimate the upper limit of bottom contribution, we
use the lower limit of the decay electron spectrum, shown as the
open circles. B-meson spectrum (solid curve) and its decay
electron spectrum (open triangles) are normalized by varying the
ratio of $\sigma_{b\bar{b}}/\sigma_{c\bar{c}}$. The summed
spectrum (solid circles) by combining the lower limit of
$D\rightarrow e$ and $B\rightarrow e$ is used to fit STAR data in
\pp\ collisions, and then the upper limit of $B\rightarrow e$
contribution will be extracted.

\begin{figure} \centering \begin{minipage}[c]{0.47\textwidth} \centering
\includegraphics[width=1.\textwidth]{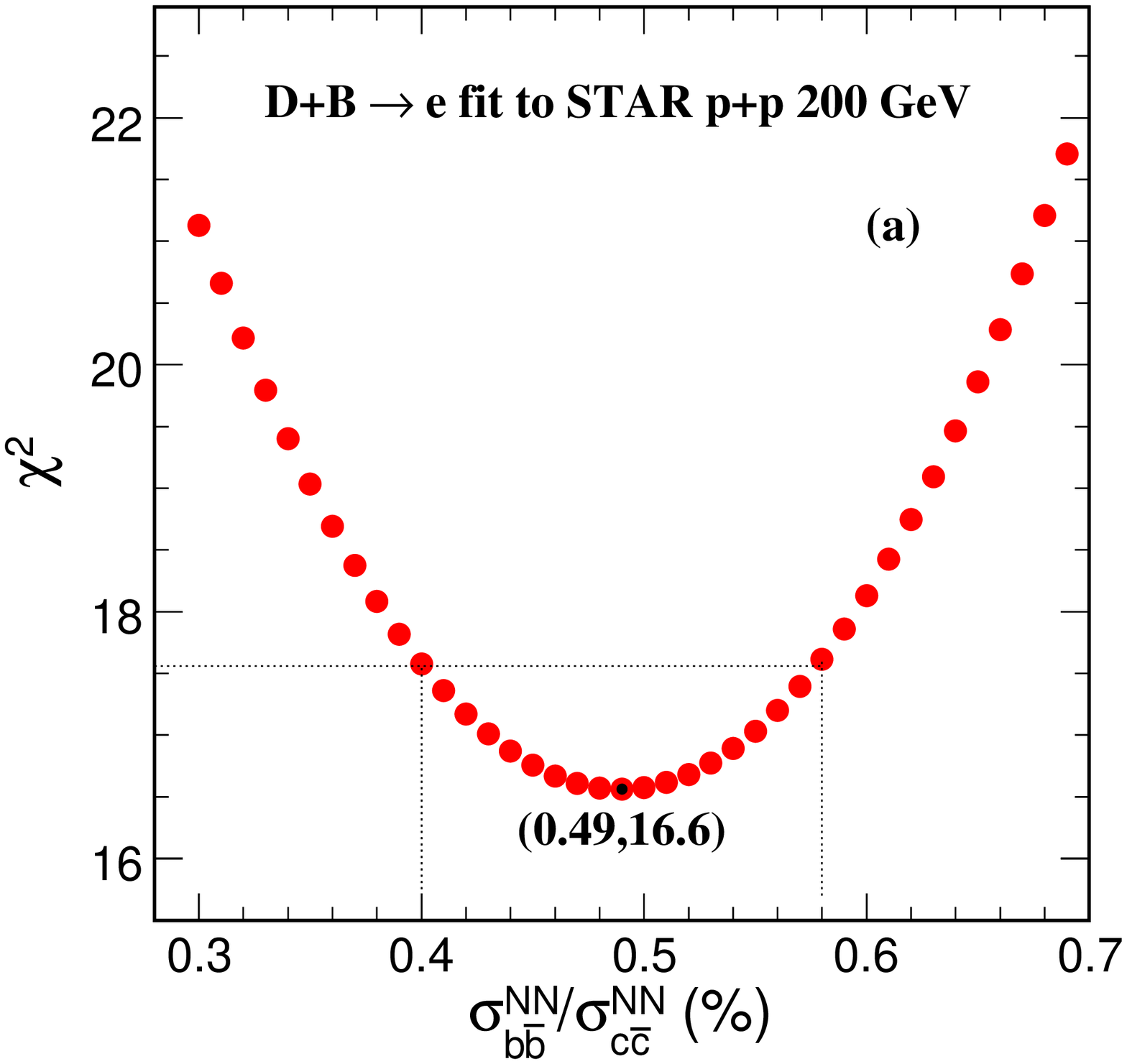}
\end{minipage}%
\begin{minipage}[c]{0.47\textwidth} \centering
\includegraphics[width=1.\textwidth]{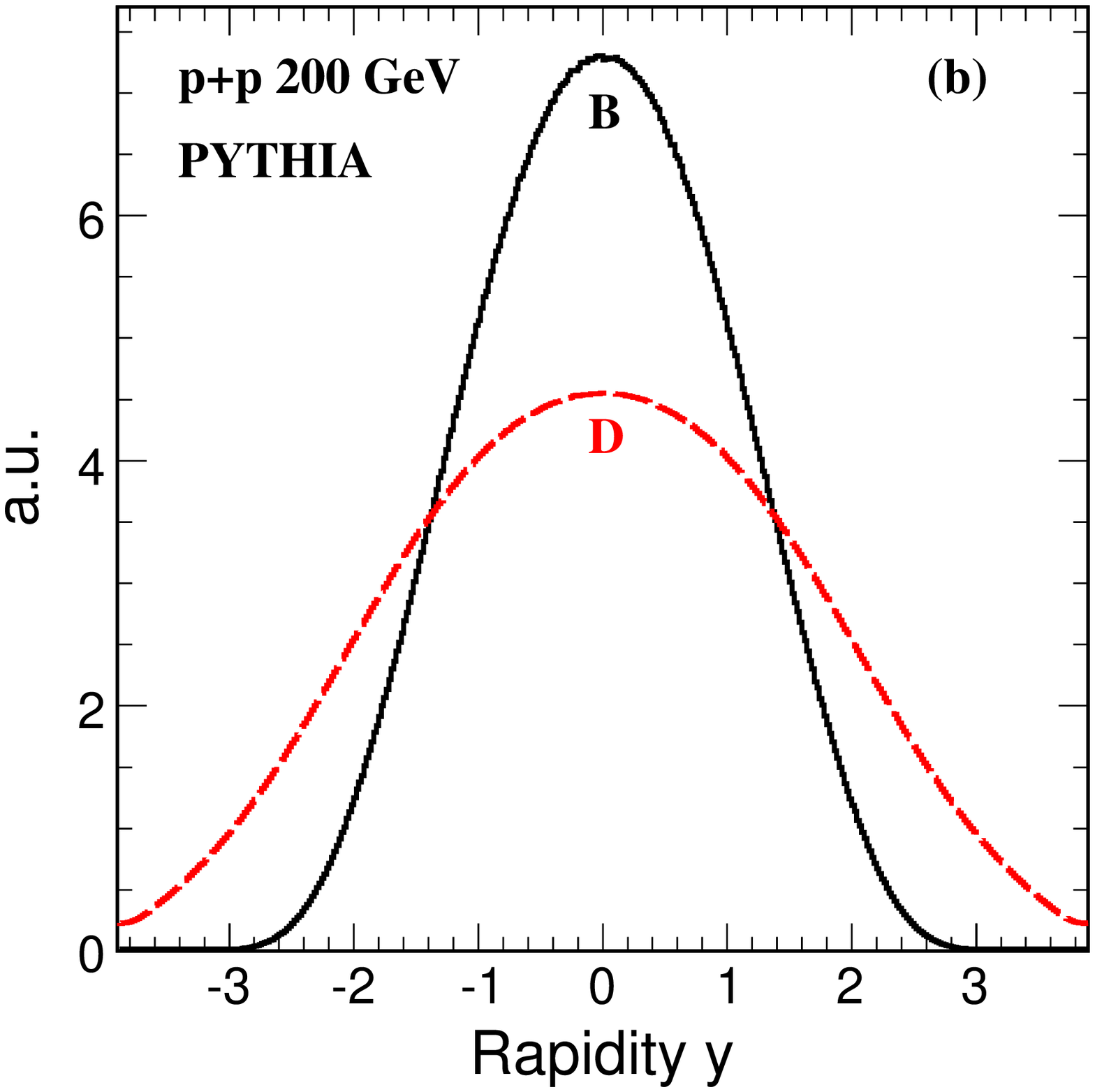}
\end{minipage}%
\caption[$\chi^2$ distribution from fitting to non-photonic
electron spectrum and rapidity distributions from PYTHIA]{Panel
(a): Fit $\chi^2$ as a function of
$\sigma_{b\bar{b}}/\sigma_{c\bar{c}}$. Straight lines is for the
$\chi^2=\chi_{min}^2+1$. Panel (b): B- (solid curve) /D- (dashed
curve) mesons rapidity distributions from PYTHIA.}
\label{pyfitchi2} \end{figure}

Fig.~\ref{pyfitchi2} (a) shows the fit $\chi^2$ as a function of
the unique variable $\sigma_{b\bar{b}}/\sigma_{c\bar{c}}$. The
best fit with a minimum $\chi^2/ndf=16.6/14$ gives the upper limit
of the total cross section ratio as
$\sigma_{b\bar{b}}/\sigma_{c\bar{c}}\leq(0.49\pm0.09\pm0.09)\%$.
The first term of the errors is calculated from
$\chi^2=\chi_{min}^2+1$. The second term is from the 15\%
normalization error of the $dN/dy$ converted to total cross
sections due to the uncertainties of the model dependent rapidity
distributions~\cite{Xinthesis}. Fig.~\ref{pyfitchi2} (b) shows the
B-/D- mesons rapidity distributions from PYTHIA. The cross section
ratio from FONLL calculation is 0.18\%-2.6\%~\cite{cacciari}. The
upper limit is consistent with theory prediction.

The upper limit of $B\rightarrow e$ contributions as a function of
\pt\ is shown in Fig.~\ref{pyspecratio} (b). It is increasing and
becomes flat around 7 \gevc. The \pt\ crossing point, where the
bottom contribution is equal to charm, of electron spectra from
B,D decay is very sensitive to the cross section ratio, since at
high \pt, these electron spectra shapes are similar. From the
$B+D\rightarrow e$ fit to STAR \pp\ data, we estimate the crossing
point $p_T^c\geq7$ \gevc.

Table~\ref{crosspt} lists the crossing points of heavy flavor
decay electrons in several \pt\ bins.

\begin{table}[hbt]
\caption[Crossing points of heavy flavor decay electrons]{Crossing
points of heavy flavor decay electrons as a function of $p_T$.}
\label{crosspt} \vskip 0.1 in
\centering\begin{tabular}{|c|c|c|c|c|c|c|c|} \hline \hline
\pt\ (\gevc) & 2 & 3 & 4 & 5 & 6 & 7 ($p_T^c$) & 8 \\
\hline
$(B\rightarrow e)/(D\rightarrow e)\leq$ & 0.11 & 0.31 & 0.53 & 0.77 & 0.85 & 1.2 & 1.1 \\
\hline \hline
\end{tabular}
\end{table}

\section{Fit to non-photonic electron $v_2$}\label{others}

Besides the non-photonic electron spectrum, the non-photonic
electron $v_2$ has also been measured in 200 \gev\ \AuAu\
collisions at RHIC~\cite{Phenixv2}. In this measurement, bottom
contribution has not been separated, which can be studied by
comparing simulations and data. Since heavy flavor hadrons \pt\
distributions and $v_2$ are unknown, our simulations have to base
on the following assumptions:

\begin{description}
\item[--] The same relative $(B\rightarrow e)/(D\rightarrow e)$
ratio from \pp\ to \AuAu. \item[--] Assume the B-/D- meson $v_2$
as the inputs for the simulation, here we assume three aspects:
\begin{itemize}
\item \uppercase \expandafter {\romannumeral 1}: B-/D- meson $v_2$
are similar as light meson $v_2$. \item \uppercase \expandafter
{\romannumeral 2}: D-meson $v_2$ as light meson $v_2$ but B-meson
does not flow. \item \uppercase \expandafter {\romannumeral 3}:
$B\rightarrow e$ contribution is neglected and D-meson $v_2$
decreases at $p_T>2$ \gevc.
\end{itemize}
\end{description}

Here heavy flavor baryons, $\Lambda_c$, $\Lambda_b$ are taken into
account as 10\% of total heavy flavor hadrons~\cite{PDG,pdgerr}.
Their $v_2$ are assumed to follow light baryon $v_2$. This baryon
contribution effect in this simulation is small.

\begin{figure} \centering\mbox{
\includegraphics[width=0.9\textwidth]{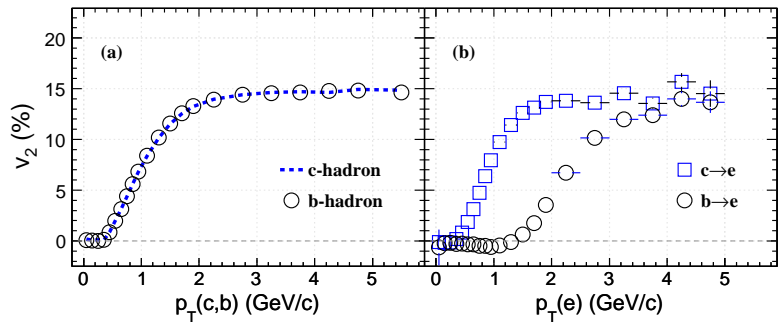}}
\caption[Decay electron $v_2$ from assumed B-/D- mesons
$v_2$]{Panel (a): Assumed B-meson $v_2$ (open circles) and D-meson
$v_2$ (dashed curve) as light meson $v^2$. Panel (b): Electron
$v_2$ from B-meson decays (open circles) and D-meson decays (open
squares).} \label{cbev2} \end{figure}

We use the light meson $v_2$ curve from fitting experimental
data~\cite{minepiv2} as the input B/D $v_2$ distributions
(Assumption \uppercase \expandafter {\romannumeral 1}), see
Fig.~\ref{cbev2} (a). That means in each \pt\ bin, the B/D
$\Delta\phi$ distribution is initialized. The electron
$\Delta\phi$ distributions in each \pt\ bin will be obtained via
B/D decays in PYTHIA model. Then the electron $v_2$, shown in
Fig.~\ref{cbev2} (b), will be extracted by fitting the
$\Delta\phi$ distributions in each \pt\ bin.

Fig.~\ref{cbev2} shows the obvious mass effect: The B/D $v_2$ are
assumed as the same, but the decay electron $v_2$ can be very
different due to decay kinematics. This is not surprising, since
we know B-meson is much heavier than D-meson and light hadrons.
The decay electrons can only have a small momentum fraction of
B-mesons. The momentum and angular correlations between decay
electrons and B-mesons are weak, especially at low \pt. Therefore,
at low \pt\ the decay electron $\phi$ angle will almost randomly
distribute. So we see the zero or negative $v_2$ for the electron
from B-meson decays. But from previous study, we know that bottom
contribution below 3 \gevc\ is small, thus the mass effect to the
total electron $v_2$ is not significant.

\begin{figure} \centering\mbox{
\includegraphics[width=0.9\textwidth]{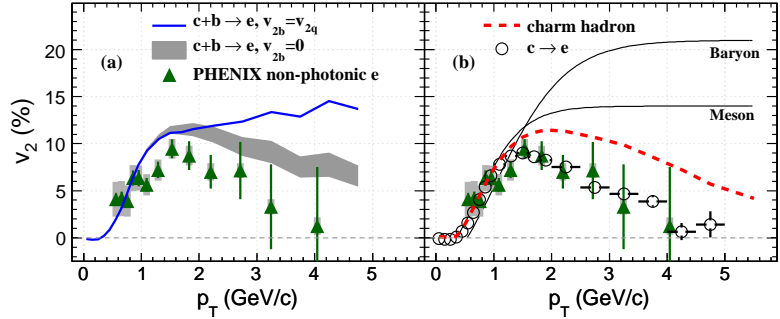}}
\caption[The total electron $v_2$ from PYTHIA simulation compared
to data]{Panel (a): The total electron $v_2$ from PYTHIA
simulation assuming that bottom flows (solid curve) and bottom
does not flow (band) compared to data. Panel (b): The total
electron $v_2$ from PYTHIA simulation fit to data and the
estimated D-meson $v_2$.} \label{v2bnob} \end{figure}

Fig.~\ref{v2bnob} (a) shows the total electron $v_2$ from PYTHIA
simulation compared to data. The measured non-photonic electron
$v_2$ from PHENIX is shown as the triangles. The solid curve
(Assumption \uppercase \expandafter {\romannumeral 1}) is the sum
$v_2$ of the two decay electron $v_2$ distributions in
Fig.~\ref{cbev2} (b) by taking the relative ratio of
$(B\rightarrow e)/(D\rightarrow e)$ into account. It can not
describe the data. If we assume B-meson does not flow (Assumption
\uppercase \expandafter {\romannumeral 2}), the total decay
electron $v_2$ will become decreasing, shown as the band. The band
is corresponding to the
$\sigma_{b\bar{b}}/\sigma_{c\bar{c}}=(0.3-0.7)\%$ (The upper
limit, 0.49\%, is in between). It has better agreement with data,
but still higher. The decreasing of non-photonic electron $v_2$
could be due to $B\rightarrow e$ contribution and B-meson $v_2$
could be very small. But below 3 \gevc, $B\rightarrow e$
contribution is not significant. That indicates D-meson $v_2$
should be smaller than light meson $v_2$ and start decreasing at
higher \pt\ ($>2$ \gevc).

So ignoring $B\rightarrow e$ contribution, we try to speculate the
D-meson $v_2$ by fitting the data using decay electron $v_2$
(Assumption \uppercase \expandafter {\romannumeral 3}). In
Fig.~\ref{v2bnob} (b), the best fit of the decay electron $v_2$ is
shown as the open circles. The estimated D-meson $v_2$ is shown as
the dashed curve, which is smaller than light meson $v_2$ above 1
\gevc\ and start decreasing above 2 \gevc.

\section{Conclusions}\label{concl}

Charm/bottom and their decayed electron spectra and $v_2$ have
been studied using PYTHIA simulation. From fitting to the STAR
non-photonic electron spectra in p+p collisions, we estimate the
upper limit of the total cross-section ratio as
$\sigma_{b\bar{b}}/\sigma_{c\bar{c}}\leq(0.49\pm0.09\pm0.09)\%$.
And the crossing point of electron spectra from B decay and D
decay is estimated as $p_T^c\geq7$ \gevc.

The bottom contribution due to mass effect can decrease the
non-photonic electron $v_2$, but this effect is not significant.
The decrease of the non-photonic electron $v_2$ is mainly due to
the decrease of the parent D-meson $v_2$. The estimated D-meson
$v_2$ is smaller than light meson $v_2$ above 1 \gevc\ and start
decreasing above 2 \gevc. This most possible D-meson $v_2$
distribution shows that at $p_T<3$ \gevc, where the bottom
contribution is negligible, D-meson has large $v_2$, indicating
that charm strongly flows in high dense medium, which could be the
evidence of light flavor thermalization in QGP created at RHIC
energy.

\section*{Acknowledgments}
We thank to S.~Esumi, H.~Huang, Y.~Miake, S.~Sakai and N.~Xu for
their cooperation. We thank to the conference organizers. We would
also like to appreciate Drs. L.J.~Ruan and Z.B.~Xu for helpful
discussions.

\vfill\eject

\begin{thebibliography}{1}
\bibitem{bib9}
From Eq.~(\ref{eq2}) one sees that in the original frame the cross term is
  given by: $$ Q_{\perp L}^2 = \left( <x\;\tau\;{\rm sh}\eta'>- <x><\tau\;{\rm
  sh}\eta'>\right) $$ up to small logarithmic corrections. This clearly
  vanishes as soon as the source becomes reflection symmetric under $\eta' = -
  \eta'$.
\end{thebibliography}
\end{document}